\documentclass[epj]{svjour}

\usepackage[dvips]{graphicx}
\usepackage[latin1]{inputenc}
\usepackage{amssymb}
\usepackage{amsmath}

\begin{document}

\title{Non-Amontons behavior of friction in single contacts}

\author{L. Bureau \thanks{\email{lionel.bureau@insp.jussieu.fr}}, T. Baumberger \and C. Caroli}
\date{\today}

\institute{Institut des Nanosciences de Paris, UMR 7588 CNRS-Universit\'e Paris 6, 140 rue de Lourmel, 75015 Paris}

\abstract{We report on the frictional properties of a single contact between a glassy polymer lens and a flat silica substrate covered either by a disordered
or by a self-assembled alkylsilane monolayer. We find that, in contrast to common belief, the Amontons proportionality between frictional and normal
stresses does not hold. Besides, we observe that the velocity dependence of the sliding stress is strongly sensitive to the structure of the silane layer. Analysis of the
frictional rheology observed on both disordered and self-assembled monolayers suggests that dissipation is controlled by the plasticity of a glass-like
interfacial layer in the former case, and by pinning of polymer chains on the substrate in the latter one.
}

\maketitle

\section{Introduction}
\label{sec:intro}

Dry friction between macroscopic ``hard'' solids commonly involves multicontact interfaces, {\it i.e.} interfaces comprised of a set of micrometer-sized contacts 
between the asperities of the rough surfaces. In order to understand the physical mechanisms responsible for frictional dissipation under such 
conditions, recent experimental studies of static 
and low-velocity friction
($V\leq 100\,\mu$m.s${^-1}$, {\it i.e.} negligible
self-heating) have been performed. Two important features emerge from these studies\cite{epje}: 

(i) at constant real contact area between the solids, the static threshold stress slowly 
increases with the time spent at rest, and its growth rate increases with the shear stress applied during ageing, 

(ii) the steady-state sliding stress systematically exhibits, in the upper part of the velocity range investigated (typically 1--100 $\mu$m.s$^{-1}$), a regime of 
quasi-logarithmic {\it increase} with velocity. 

This phenomenology has been shown to be consistent with the following picture, akin
to that of ageing and plasticity of glassy media: frictional dissipation localizes in an amorphous interfacial layer of nanometric thickness, 
adhesively pinned to the substrates. At rest, this layer is the seat of a glass-like relaxation process that gives rise to static ageing. Sliding rejuvenates the layer, and 
dissipation occurs by thermally assisted depinning and rearrangement of structural units of volume $\sim$ nm$^3$, leading to the observed velocity-strengthening friction at 
velocities large enough for rejuvenation to be fully effective .

\medskip

On the other hand, many tribological studies rely on the surface force apparatus (SFA) technique, which involves a well controlled {\it single} contact configuration. 
They mostly focus on the 
frictional behavior of fluids solidified under the effet of confinement down to molecular thicknesses\cite{drummond,gourdon}, and, recently, on dry friction
between thin films of glassy polymers\cite{luengo,tirrell1}. In the former systems, frictional dissipation takes place in the nanometer-thick confined layer. In the latter ones, it involves molecular 
 rearrangements in a nanometer-thick interfacial region between the polymer films, as in the multicontact 
situation above. 

The behaviors observed in SFA and multicontact experiments do share some qualitative features, {\it e.g.} static ageing. However, several 
contrasting results emerge: 
in particular, SFA studies always evidence unstable (stick-slip) sliding 
up to a system-dependent critical velocity $V_{c}$ in the range $\sim 0.1$--$10\,\mu$m.s$^{-1}$. For $V>V_{c}$, friction is always found to be {\it velocity-weakening}.
 This is one of the most striking differences between
SFA and multicontact experiments, and two questions arise from it: 

(i) To what extent is the sliding dynamics sensitive to the chemical nature of the confined medium on the one hand, and to the strength of its interactions with the 
confining surfaces on the other hand? 

(ii) Does the level of confining pressure affect the dissipative mechanisms, and if so, how?

The latter question is of particular importance. Indeed, in SFA experiments, the confining pressure can be varied typically up to, at most, 10 MPa. 
This contrasts with the situation at multicontact interfaces, where pressure is self-adjusted and considerably higher: due to the randomness of surface profiles, 
the number of microcontacts adjusts so that, at 
essentially all usual apparent pressure levels,  
the pressure exerted on the microcontacts 
reaches a quasi-constant level close to the yield stress of the material\cite{greenwood}. For a glassy polymer like poly(methylmethacrylate) (PMMA), this pressure
is $\sim$300 MPa, {\it i.e.} one to two orders of magnitude higher than the pressure in a SFA. This in turn raises the following question: is it legitimate to ``extrapolate'' 
the results obtained in SFA 
experiments to macroscopic situations where the pressure is much higher ?

Briscoe and Tabor already adressed, in a pioneer work, the question of the pressure dependence of polymer friction\cite{tabor}. They studied the shear behavior of 
glassy polymer thin 
films, confined between a flat glass substrate and a spherical glass slider. They claimed that their data supported the existence of an Amontons-like linear dependence of the
sliding stress on the contact pressure, a statement which has become common wisdom. However, as will be discussed below, close inspection of their 
results gives no clear evidence for such a linear relationship. We believe that their work suffered from two main drawbacks: poor control of the physico-chemical state of the 
glass surfaces used, and lack of measurement of the contact area between the solids (their estimate of shear and normal stresses, based on the assumption of a non-adhesive
 hertzian contact,
may thus be biased).

For these reasons,
we have revisited the question of the influence of pressure on dynamic friction, taking care to avoid the problems that 
render the conclusions of Briscoe and Tabor questionable. We report, in this article, on friction experiments performed between a PMMA lens and a 
flat silica substrate on which an alkylsilane layer is chemically grafted. We have developed an experimental setup similar to that of Vorvolakos and 
Chaudhury\cite{chaudhury}, 
which allows for direct optical measurement of the contact area. This sphere/flat single contact configuration 
enables us to work at applied pressures in the range 10--70 MPa, intermediate between
SFA and multicontact levels.

\medskip

Under these conditions, we show that,
for PMMA sliding on a disordered layer of trimethylsilane (TMS), {\it i.e.} the same interface as in previous multicontact experiments\cite{epje}:

(i) In contrast with the conclusions of Briscoe and Tabor, the shear stress $\sigma$, measured in steady sliding at constant velocity, does not increase linearly with pressure. 
The increase of $\sigma$ with $p$ is found to be strongly sublinear in the range  of pressure 10--70 MPa.  Moreover, extrapolating this nonlinear
dependence up to $p\simeq 300$ MPa leads to a shear stress compatible with the friction coefficient measured in the multicontact configuration.

(ii) The shear stress $\sigma$, which velocity dependence has been fully investigated at two different pressures,  is found to increase logarithmically with velocity 
over the range 0.1--100 $\mu$m.s$^{-1}$. Analysis of these data leads to conclude that
the activation volume characteristic of elementary dissipative processes at such an interface is essentially insensitive to pressure in the range 35---300 MPa.

For PMMA sliding on a self assembled monolayer of octadecyltrichlorosilane (OTS), on which the friction level is about ten times lower than on TMS, we find that:

(i) the shear stress exhibits a fonctional dependence on pressure similar to that observed on TMS, 

(ii) the lower the pressure, the more strongly $\sigma(V)$ departs from simple logarithmic strengthening. At low enough pressures, a velocity-weakening regime, associated with stick-slip,
appears at low velocities.

Comparison of the behaviors on both types of silane layers suggests the following: the strong adhesive sites on which the polymer molecules get pinned onto the substrate are
scarcer and more difficult to access on the self-assembled (OTS) than on the disordered (TMS) layer. The time scale for pinning, much longer on the OTS-covered
substrate, might thus become relevant in the sliding dynamics, as in Schallamach's model of friction\cite{Schalla}. This could account for the observed stick-slip and velocity-weakening at
low pressures.

\section{Experiments}
\label{sec:exp}

\subsection{Experimental setup}
\label{subsec:setup}

\begin{figure}[htbp]
$$
\includegraphics[scale=0.5]{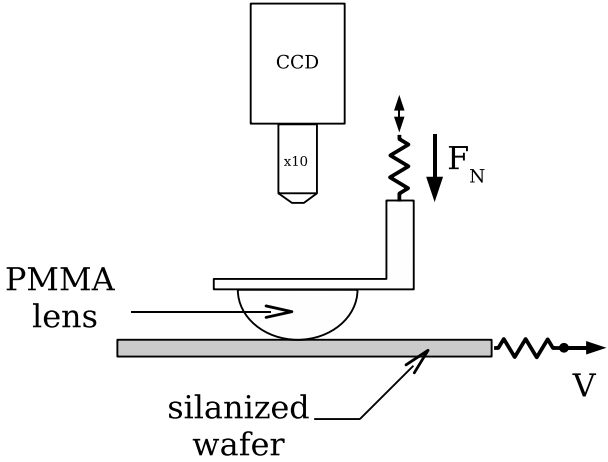}
$$
\caption{Experimental setup: a PMMA lens is pressed against a silanized substrate under a constant normal load. The substrate is pulled at velocity $V$ through a 
spring of stiffness $K_{T}$. The contact area is monitored optically.}
\label{fig:setup}
\end{figure}

The experimental setup is sketched on Figure  \ref{fig:setup}. A lens of poly(methylmethacrylate) is fixed on a transparent holder attached to one end of
a double cantilever spring (stiffness $K_{N}=2\times 10^4$ N.m$^{-1}$), the other end of which is fixed on a motorized translation stage. The lens is pressed 
against a flat horizontal substrate, and the value of the applied normal force is deduced from the spring deflection, measured by means of a capacitive displacement gauge. The range of
accessible normal forces $F_{N}$ is 4$\times$10$^{-3}$---3 N. The silicon wafer used as a substrate is attached to a double cantilever spring 
(stiffness $K_{T}=1.7\times 10^{4}$ N.m$^{-1}$), the free end of which is driven at constant velocity $V$ in the range 10$^{-1}$---10$^{2}$ $\mu$m.s$^{-1}$ by a horizontal translation stage.
The resulting tangential force $F_{T}$ is measured to within 5$\times$10$^{-4}$ N by means of a capacitive sensor. In order to work at constant normal load
during sliding, and to compensate 
for parallelism defects of the mechanical setup, a feedback loop controls the position of the vertical stage which drives the loading spring. 

The contact area $A$ between the 
polymer lens and the wafer is observed in reflexion by means of a long working distance optics and a computer-interfaced CCD camera. 
The lenses have radii of curvature on the order of a millimeter (see section \ref{subsec:samples} below), which --- for $F_{N}$ in the range given above --- 
yields contact areas typically ranging from 3$\times 10^{2}$ to 3$\times 10^{4}$ $\mu$m$^{2}$. Contact areas are determined with a $\pm 2\%$ accuracy by image processing.

The whole experimental setup is enclosed in a glovebox purged with dry argon.

\subsection{Samples}
\label{subsec:samples}

The substrates are 2'' silicon wafers covered by a silane layer. 
The wafers are cleaned as follows: rinsing with toluene, drying in nitrogen flux, 
15 minutes of sonication in a dilute solution of RBS detergent in deionized water, 15 minutes of sonication in ultra-pure water, drying in nitrogen flux, 30 minutes in 
a UV/O$_{3}$ chamber. Two different types of alkylsilanes are employed for surface modification:\\
(i) A trimethylsilane (TMS), grafted by exposition of the clean wafer to the vapor of 1,1,1,3,3,3 hexamethyldisilazane (Sigma-Aldrich) at T=80$^{\circ}$C and p$\simeq 1$ mbar for
120 hours. TMS does not self-assemble, and the disordered monolayers thus formed are known to have a thickness\cite{sugimura} of about 5 \AA.\\
(ii) Octadecyltrichlorosilane (OTS, Sigma-Aldrich). We follow a grafting protocole akin to that described by Silberzan {\it et al.}\cite{silberzan} 
and Davidovits {\it et al.}\cite{goldmann}: the wafers are exposed 
 to a flux of humid oxygen for 2 minutes immediately after UV/O$_{3}$, and are then immersed in a solution composed of 70 ml of hexadecane, 15 ml of carbon tetrachloride, 
 200 $\mu$l of OTS. Wafers are left for 5 minutes in this reaction bath at 18$^{\circ}$C, then rinsed with carbon tetrachloride. All the reagents are anhydrous grade 
 (Sigma-Aldrich) and used as received. The reaction is conducted in a glovebag under dry nitrogen. Under such conditions, the thickness of the OTS layer, measured by 
 ellipsometry, is 21$\pm 1$ \AA, in agreement with values previously reported for the same type of self-assembled monolayers\cite{silberzan}.
 
 The polymer lenses are made as follows: about 10 mm$^{3}$ of poly(methylmethacrylate) (PMMA) powder (${M_{w}}$=93 kg.mol$^{-1}$, $M_{n}=46$
 kg.mol$^{-1}$, $T_{g}\simeq 100^{\circ}$C,
 from Sigma-Aldrich) 
 is brought to
 T=250$^{\circ}$C at p=10$^{-1}$ mbar until a clear and homogeneous melt is obtained. The melt is then transferred on a clean glass slide and allowed to spread at 
 T=200$^{\circ}$C and atmospheric pressure. During the first minutes of spreading, the highly viscous polymer melt forms a spherical cap, which radius of
 curvature increases with the spreading time. Once the spherical cap has reached a roughly millimetric radius of curvature, the melt and its glass holder are transferred into
  an oven at 
 80$^{\circ}$C and left at this annealing temperature for 12 hours.
 
  The curvature of the lenses is deduced from the radius of the 
 Newton rings that form when the lens, brought close to contact with the reflective substrate, is illuminated with monochromatic light. 
 
\begin{figure}[htbp]
$$
\includegraphics[scale=1]{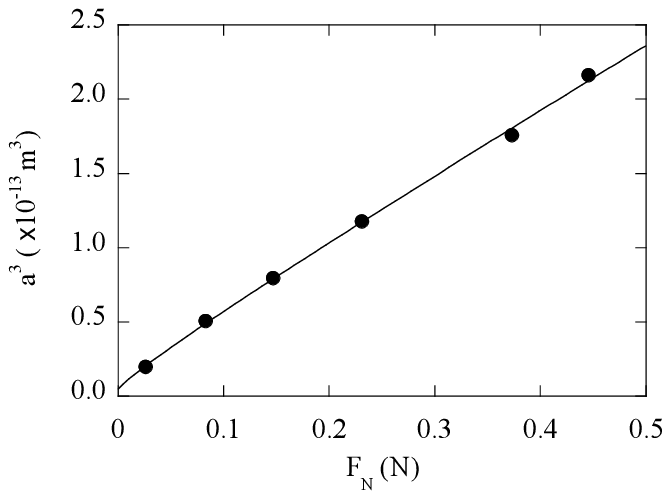}
$$
\caption{JKR plot of  $a^3$ ($a$ is the contact radius) as a function of $F_{N}$, for a PMMA lens of radius of curvature $R=2$ mm, pressed against a 
TMS-covered substrate. No hysteresis is observed between loading and unloading. ($\bullet$): experimental data; (---): best JKR fit yielding a Young modulus $E=2.9$
MPa and an adhesion energy $\gamma=0.3$ J.m$^{-2}$.}
\label{fig:hertz}
\end{figure}

 In order to check the elastic properties of the PMMA samples, 
 we put each lens in contact 
 with the  substrate and measure $a$, the radius of the circular contact area, for various normal loads $F_{N}$. For all samples, we find a linear increase of $a^3$ with $F_{N}$ (see 
 Figure \ref{fig:hertz}), with a positive offset at zero applied load due to adhesion. A fit of these data according to the JKR theory\cite{JKR}  
  yields the Young modulus of the lens, 
 which is found to lie in the range 3---3.6 GPa, in good agreement with what has been measured in uniaxial compression of bulk PMMA\cite{prb1}. For TMS substrates,
 the adhesion energy deduced from the fit is
 found to be in the range 0.3---0.5 J.m$^{-2}$, {\it i.e.} about five times higher than what we could expect for the contact of PMMA on a methyl-terminated surface. 
 This indicates that the TMS layers probably exhibit coverage defects, and that
  hydrogen bonding occurs between PMMA molecules and free Si-OH groups on the underlying silicon oxide surface.
 
\section{Pressure dependence of the friction stress}
\label{sec:sigp}

\begin{figure}[htbp]
$$
\includegraphics[scale=1]{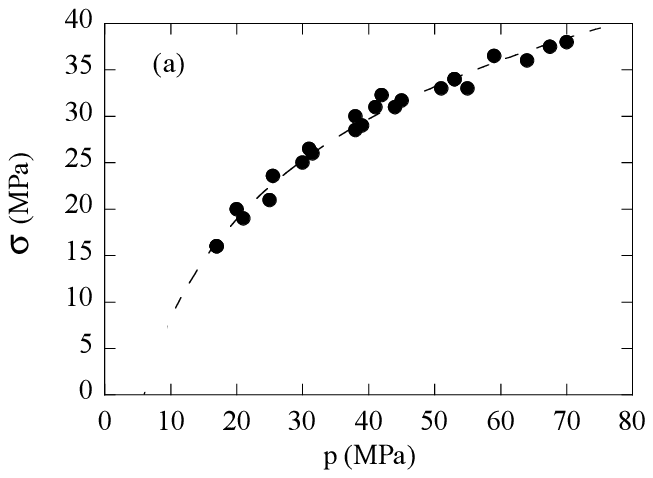}
$$
$$
\includegraphics[scale=1]{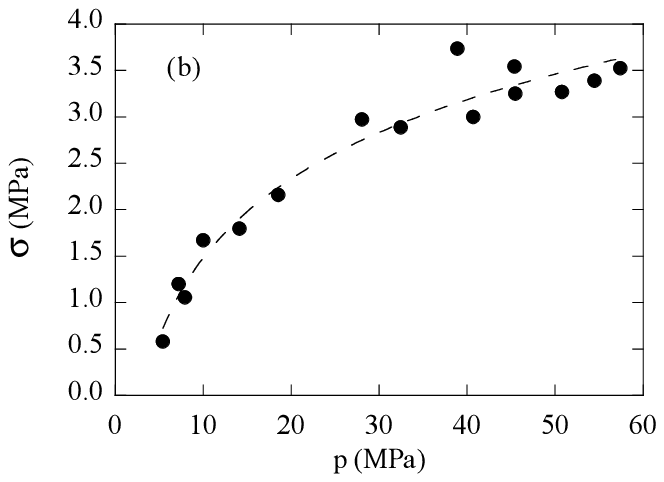}
$$
\caption{Pressure dependence of the friction stress   measured
in steady sliding at V = 10 $\mu$m.s$^{-1}$. (a): PMMA sliding on
TMS. (b): PMMA sliding on OTS. The dashed lines correspond
to the best logarithmic fit of the data.}
\label{fig:sigp}
\end{figure}

\begin{figure}[htbp]
$$
\includegraphics[scale=1]{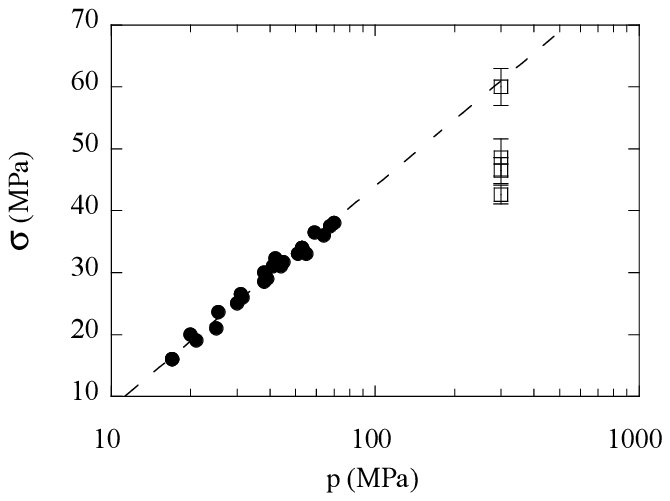}
$$
\caption{Friction stress as a function of pressure. ($\bullet$): same data
as on Figure 3(a). ($\square$) $\sigma=\mu_{d}p$ estimated from measurements
of $\mu_{d}(V=10\, \mu\text{m.s}^{-1})$ at PMMA/TMS multicontact interfaces.
The error bars account for observed fluctuations of $\mu_{d}$ while
sliding along the same track. The dashed line is the best loga-
rithmic fit to the sphere/flat data.}
\label{fig:sigpextra}
\end{figure}

In a first set of experiments, we have measured the average sliding stress $\sigma=F_{T}/A$ at the constant velocity $V=10\, \mu$m.s$^{-1}$ at various average normal
stresses $p=F_{N}/A$. In the following, we will simply call them ``friction stress'' and ``pressure''. The results are shown on figures \ref{fig:sigp}a (TMS substrate) and 
\ref{fig:sigp}b (OTS substrate). Note that, although the friction level on OTS is about one order of magnitude lower than on TMS,
in both cases, the growth of $\sigma(p)$ is clearly strongly sublinear, and, as seen on figures \ref{fig:sigp} and \ref{fig:sigpextra}, the two sets of data are  
well fitted 
by an empirical logarithmic law 
over the pressure range 10--70 MPa. 

On the other hand, we previously measured the dynamic friction coefficient $\mu_{d}(V)=F_{T}(V)/F_{N}$ for a multicontact interface between rough PMMA and a 
flat glass substrate covered by TMS under the same protocole as described in section \ref{sec:exp}\cite{epje}. From these data we estimate the sliding stress as $\sigma=\mu_{d}p$,
where $\mu_{d}$ is measured at $V=10\, \mu$m.s$^{-1}$ and $p$ is the average normal stress over the load-bearing microcontacts. We take for the value of $p$ the 
hardness of PMMA, $H=300$ MPa, an upper limit corresponding
to fully plastic asperity deformation. This provides us with data at a pressure much higher than the maximum value reached with the sphere/flat setup. It is seen on Figure 
\ref{fig:sigpextra} 
that the logarithmic fit to the sphere/flat data, when extrapolated to high pressures, gives an estimate which, though somewhat higher, is compatible with the
 multicontact results.

This brings further support to our conclusion, namely that Amontons-Coulomb proportionality between frictional and normal forces, while valid for multicontact 
interfaces, does not hold on the local level of single contacts. For these, the sliding stress grows with pressure in a strongly sublinear fashion.

\begin{figure}[htbp]
$$
\includegraphics[scale=1]{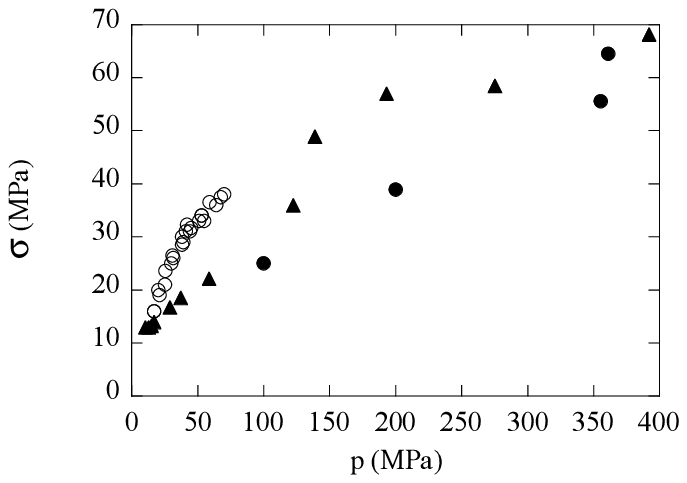}
$$
\caption{Sliding stress as a function of pressure. ($\circ$): same data
as on Figure 3(a). ($\bullet$): data taken from Fig. 2 of reference \cite{tabor}.
($\blacktriangle$): data taken from Fig. 1 of reference \cite{tabor}.}
\label{fig:tabor}
\end{figure}

The contradiction betweeen this conclusion and that previously claimed by Briscoe and Tabor (BT)\cite{tabor} has led us to reexamine their data, which we replot on Figure 
\ref{fig:tabor}\cite{note1}, along
with our PMMA/TMS results. The full dots are those of their data from which they concluded to linearity, while the full triangles, transcribed from their Figure 1, 
correspond to another set of data
pertaining to the same PMMA/glass interface. Consideration of the full set of results obviously casts doubt on their conclusion.

\begin{figure}[htbp]
$$
\includegraphics[scale=1]{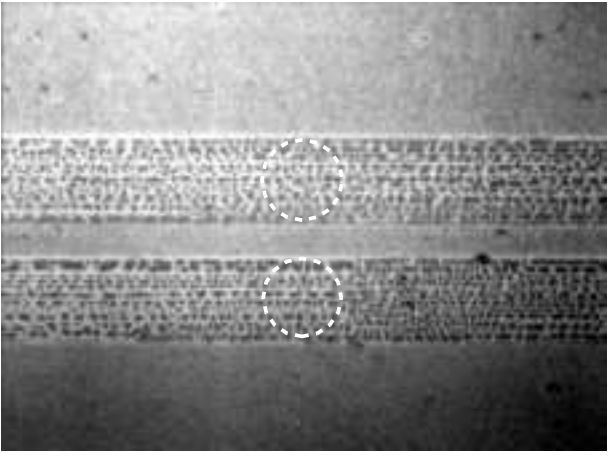}
$$
\caption{Breath figure formed on exposing to water-saturated
air a bare glass surface onto which a PMMA lens has slid along
two parallel tracks. A continuous wetting film forms over clean
glass, whereas a droplet pattern decorates the two tracks. The
dashed circles indicate the size of the contact zone for both
tracks. Image size is 615$\times$455 $\mu$m$^{2}$.}
\label{fig:souffle}
\end{figure}

Now, it must be kept in mind that, in their experiments, PMMA was sliding upon nominally bare glass. Multicontact experiments have shown that the friction level of PMMA
is much higher on bare glass than on silanized substrates. It is therefore surprising that most of the BT data lie below ours. This strongly suggests that, due to strong adhesion,
some PMMA chains get transferred onto the glass substrate, a point which we have checked as follows: after having slid a PMMA sphere along clean glass, we have exposed the glass flat 
to water-saturated air. The resulting breath figure\cite{souffle} unambiguously reveals the non-wettability of the sliding trace (see Fig. \ref{fig:souffle}), hence the 
presence of PMMA. We are then lead to
attributing the scatter of the BT data to the gradual build-up of a transferred PMMA film as the number of traversals over the same track increases, a process which we 
have minimized here by using silanized surfaces.

\section{Evolution with pressure of the frictional rheology}
\label{sec:sigV}

Comparing, as was done in the previous section, friction stresses at an arbitrary common velocity only provides a partial characterization of the pressure effect.
Indeed, it is well known that sliding stresses are velocity-dependent, and that the corresponding rheology gives insight into the nature of the underlying dissipative
mechanisms. We have therefore measured $\sigma(V)$ at several pressures, over three decades of velocity, for the two substrates. Since the results exhibit qualitative 
differences, we present them separatly.

\subsection{TMS substrate}
\label{subsec:tms}

\begin{figure}[htbp]
$$
\includegraphics[scale=1]{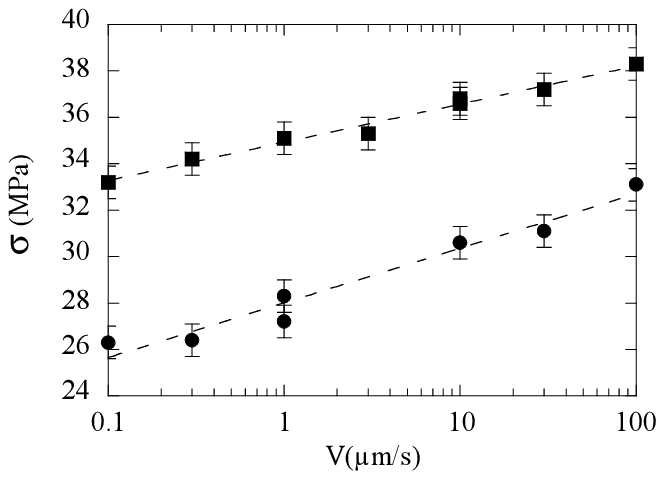}
$$
\caption{Velocity dependence of the friction stress on TMS. ($\blacksquare$):
at pressure $p = 60$ MPa. ($\bullet$): at $p = 35$ MPa. Dashed lines are
the best logarithmic fits to the data. Error bars account for
fluctuations of $\sigma$ while sliding along the same track.}
\label{fig:sigVTMS}
\end{figure}

Figure \ref{fig:sigVTMS} presents $\sigma(V)$ measured at the two pressures $p=35$ and 60 MPa. It is seen that, in the velocity range 0.1--100 $\mu$m.s$^{-1}$,
$\sigma$ increases logarithmically with $V$. Such a rheology is also observed at multicontact PMMA/TMS interfaces\cite{epje}. It can be interpreted as due to
thermally-assisted stress-induced dissipative events, corresponding to structural rearrangements of nanometric clusters within the glassy interfacial adhesive layer.

From the logarithmic slope $\alpha=d\sigma/d(\ln V)$ one may then compute an activation volume which can be considered as a measure of the average size of the clusters
involved in the elementary events: $v_{act}=k_{B}T/\alpha(p)$. We thus obtain: for p=35 MPa, $v_{act}=4.2\pm 0.5$ nm$^3$, and for p=60 MPa, 
$v_{act}=5.4\pm 0.5$ nm$^3$. For multicontact interfaces ($p\lesssim 300$ MPa)\cite{epje}, it was found that $v_{act}=2.5$--3 nm$^3$. That is, we can conclude that, for this system,
the nature of the dissipative processes is unaffected by pressure in the range 35--300  MPa. The elementary cluster size retains its order of magnitude, though showing 
a trend towards increasing with decreasing pressure. One may tentatively associate this trend with the increase of free volume in the glasslike interfacial layer as the level
 of confinement decreases.

\subsection{OTS substrate}
\label{subsec:ots}

As already mentioned, the overall friction level is about ten times lower than on TMS.  This is most probably due to the structure of the OTS grafted layer:  OTS molecules
self-assemble, which leads to higher coverage of the underlying silica surface.

\begin{figure}[htbp]
$$
\includegraphics[scale=1]{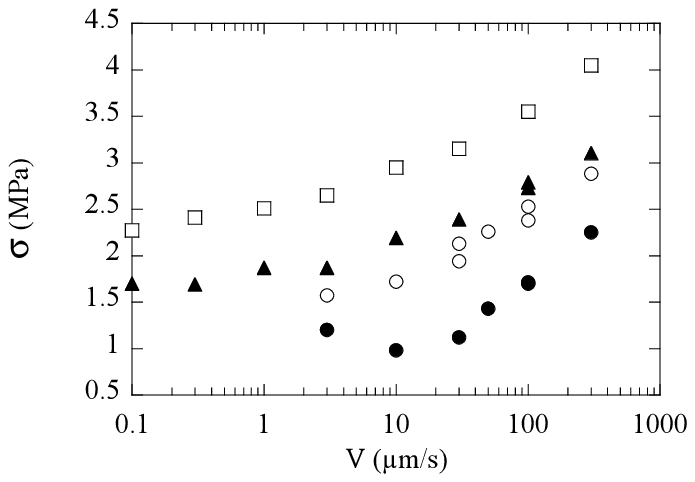}
$$
\caption{Velocity dependence of the friction stress on OTS. ($\square$):
pressure p = 55 MPa. ($\blacktriangle$): p = 30 MPa. ($\circ$): p = 15 MPa. ($\bullet$):
p = 7 MPa. Error bars are of the size of the symbols.}
\label{fig:sigVOTS}
\end{figure}

A first experiment was conducted at p=55 MPa, also revealing a velocity-strengthening rheology. However,  $\sigma (V)$ exhibits, in 
contradistinction with TMS at the same pressure, a noticeable departure from log-linearity (see Figure \ref{fig:sigVOTS}). 
This led us to investigate its behavior down to low pressures.

As appears on Figure \ref{fig:sigVOTS}, the non-linear dependence on $\ln(V)$  persists and even increases, resulting, at p=7 MPa, in the emergence of a velocity-weakening
regime below $V\simeq 10\, \mu$m.s$^{-1}$, which leads to stick-slip for $V<3\, \mu$m.s$^{-1}$. For $p=15$ MPa, stick-slip is also observed at $V\leq 1\, \mu$m.s$^{-1}$,
indicating the existence of a minimum of $\sigma (V)$ between 1 and 3 $\mu$m.s$^{-1}$.

This behavior, together with the flattening of $\sigma (V)$ at low velocities for higher pressures,
strongly suggests that a velocity-weakening regime always exists and that its upper limit $V_m$ decreases
rapidly as pressure increases, so that it lies below $0.1\, \mu$m.s$^{-1}$ at the higher pressure levels.

Conversely, this trend is consistent with the behavior observed in the low pressure ($p\sim 1$ MPa) SFA studies  of
polymer friction\cite{luengo,tirrell1}. Namely, under these latter conditions,
stick-slip and velocity-weakening friction prevail over the whole velocity range 0.01--10 $\mu$m.s$^{-1}$.

\begin{figure}[htbp]
$$
\includegraphics[scale=1]{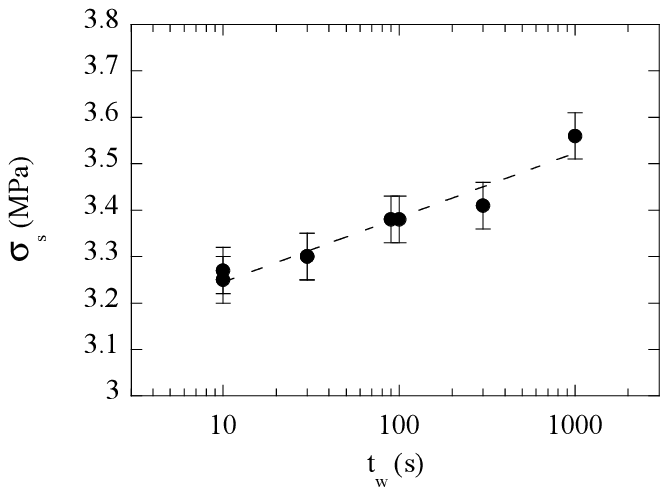}
$$
\caption{Evolution of the static friction stress  s with the time
tw spent at rest before sliding. ($\bullet$): experimental data. (---):
best logarithmic fit.}
\label{fig:ageing}
\end{figure}

A velocity-weakening regime is the signature of the existence of a structural variable, the dynamics of
which is coupled to motion. In steady motion, the value of this ``age variable'' is a measure of
the strength of the interfacial pinning. The faster the motion, the lower the pinning level, which leads to
the weakening. Such a ``rejuvenation by motion'' has its counterpart in the slow growth of the static
threshold with the time spent at rest. That static ageing is indeed at work on PMMA/OTS is illustrated on 
Figure \ref{fig:ageing}.

These observations lead us to propose the following qualitative picture. The low friction level signals 
that the OTS layer only exhibits a small fraction of coverage defects, hence a small density of 
sites available for forming strong H-bonds between PMMA surface chain segments and silica. In the spirit
of Schallamach´s model of rubber friction\cite{Schalla}, when sliding, the fraction of these bonds effectively realized
is limited by the competition between advection and the bond-formation kinetics. This leads to a 
decrease of the overall pinning level with increasing velocity, hence to a velocity-weakening contribution to
$\sigma (V)$, which becomes negligible beyond some characteristic velocity $d/\tau \sim V_m$, where $d$
 and $\tau$ are respectively the a radius and an average time of capture. One may reasonably guess that $\tau$ scales
 with the so-called $\beta$ relaxation time associated with the hindered  rotation of the -COOCH$_{3}$ side groups, which certainly helps bonding between the polar groups and the
 free silanols. 
 On the basis of their recent friction force microscopy results\cite{hammer}, Hammerschmidt {\it et al} evaluate that $\tau_{\beta}$ should
 lie in the 10$^{-4}$~s range at room temperature. With $d\sim 1$ nm, this yields a scale for $V_{m}$ indeed lying in the $\mu$m.s$^{-1}$ range.
 
Above $V_m$, the only remaining contribution to 
$\sigma (V)$ is the velocity-strengthening one, corresponding to viscous dissipation in the confined polymeric
interfacial layer. One reasonably expects that the higher the pressure, the lower the molecular mobility
in this layer, hence the longer $\tau$, in agreement with the observed decrease of $V_m$.

This picture probably does not apply to the TMS case in the investigated velocity range. For this latter 
system, the high friction level points towards a much weaker efficiency of the silanization process.
 Moreover, TMS monolayers are much thinner ($\sim 0.5 $ nm) than OTS ones ($\sim 2$ nm), making the formation
of  H-bonds both faster and more probable. On this basis, and in view of the log-linear dependence of
$\sigma (V)$, we suggest that adhesive bonding sites remain saturated up to velocities much larger than 
100 $\mu$m.s$^{-1}$, so that dissipation is completly controlled by the glasslike rheology invoked in 
section \ref{subsec:tms}.

\section{Conclusion}
\label{sec:end}

In summary, the experimental results described above lead us to state that, {\it in unlubricated single contacts
exhibiting solid friction, the Amontons proportionality between frictional and normal forces
--- hence between the corresponding stresses $\sigma$ and $p$ --- is not the rule}.

For PMMA/silane contacts, in the intermediate range lying between the pressure levels characteristic of SFA
on the one hand and multicontact on the other, the growth of $\sigma$ with $p$ is strongly sublinear.
This by no means contradicts the fact that the Amontons law holds for forces at multicontact interfaces: in this
configuration, it is a consequence of the proportionality between the real area of contact and the normal
 load $F_N$, while the average stresses on microcontacts remain load independent.

We also find that the frictional rheology may exhibit a non-trivial dependence on pressure. That is, 
studying $\sigma (p)$ at a single velocity appears as an insufficient characterization of the pressure 
effect.

Based on the evolution of $\sigma (V)$ with pressure, we have proposed two different pictures for the TMS
and OTS systems, which suggests that friction is essentially controlled by jamming in the former case, and
by pinning in the latter one. In order to test this interpretation further, and to get a handle on the
 relative weight of these two mechanisms, a more extensive study, using 
substrates with controlled partial OTS-coverage as a mean for controlling the overall level of interfacial
pinning, is presently in progress.

\medskip

We wish to thank Bastien Calmettes for his contribution to the experimental work during his stay at INSP.


\begin{thebibliography}{}
\bibitem{note1} We have excluded those of their results corresponding to an evaluated pressure higher than the hardness of PMMA. Indeed, in this regime,
 given the
smallness of the radius of curvature of the glass spheres, some plastic indendation of the PMMA film has necessarily taken place, leading to the inadequacy of their hertzian
 estimate of the pressure.
\end{thebibliography}
\end{document}